\documentclass[12pt]{iopart}

\usepackage{braket}
\usepackage[pdftex]{graphicx}
\usepackage{bm}
\usepackage{ascmac}
\usepackage{mathrsfs}
\usepackage{amsthm}
\usepackage{amssymb}

\newcommand{\eq}[1]{\begin{eqnarray} #1\end{eqnarray}}

\newcommand{\ex}[0]{{\mathrm e}}
\newcommand{\kakko}[1]{\left( #1 \right)}

\newcommand{\intd}[0]{{\rm d}}
\newcommand{\expo}[1]{\exp{\left[ #1 \right]}}
\newcommand{\pd}[3]{
 \if 1#1 \frac{\partial #2}{\partial #3}
 \else \frac{\partial^{#1} #2}{\partial #3^{#1}}\fi}
 \newcommand{\od}[3]{
 \if 1#1 \frac{{\mathrm d} #2}{{\mathrm d} #3}
 \else \frac{{\mathrm d}^{#1} #2}{{\mathrm d}#3^{#1}}\fi}

\begin{document}

\title{
Nonclassicality of open circuit QED systems in the deep-strong coupling regime
}

\author{Tomohiro Shitara$^1$, Motoaki Bamba$^{2,3,4}$, Fumiki Yoshihara$^5$, Tomoko Fuse$^5$, Sahel Ashhab$^{5,6}$, Kouichi Semba$^5$\footnote{Present address: Institute for Photon Science and Technology, The University of Tokyo, Tokyo 113-0033, Japan.}, and Kazuki Koshino$^1$}
\ead{shitara.las@tmd.ac.jp}

\address{$^1$ College of Liberal Arts and Sciences, Tokyo Medical and Dental University, 2-8-30 Konodai, Ichikawa 272-0827, Japan}

\address{$^2$The Hakubi Center for Advanced Research, Kyoto University, Yoshida-Honmachi, Sakyo-ku, Kyoto 606-8501, Japan.}

\address{$^3$PRESTO, Japan Science and Technology Agency, Kawaguchi 332-0012, Japan}

\address{$^4$Department of Physics I, Kyoto University, Kitashirakawa Oiwake-cho, Sakyo-ku, Kyoto 606-8502, Japan}

\address{$^5$National Institute of Information and Communications Technology, 4-2-1, Nukui-Kitamachi, Koganei, Tokyo 184-8795, Japan}

\address{$^6$Qatar Environment and Energy Research Institute, Hamad Bin Khalifa University, Qatar Foundation, Doha, Qatar}

\begin{abstract}
We investigate theoretically how the ground state of a qubit-resonator system in the deep-strong coupling (DSC) regime is affected by the coupling to an environment.
We employ as a variational ansatz for the ground state of the qubit-resonator-environment system a superposition of coherent states displaced in qubit-state-dependent directions.
We show that the reduced density matrix of the qubit-resonator system strongly depends on how the system is coupled to the environment, i.e., capacitive or inductive, because of the broken rotational symmetry of the eigenstates of the DSC system in the resonator phase space.
When the resonator couples to the qubit and the environment in different ways (for instance, one is inductive and the other is capacitive), the system is almost unaffected by the resonator-waveguide coupling.
In contrast, when the two couplings are of the same type (for instance, both are inductive), by increasing the resonator-waveguide coupling strength, the average number of virtual photons increases and the quantum superposition realized in the qubit-resonator entangled ground state is partially degraded.
Since the superposition becomes more fragile with increasing the qubit-resonator coupling, there exists an optimal coupling strength to maximize the nonclassicality of the qubit-resonator system.
\end{abstract}

%
%
%
\maketitle
%
%

\section{Introduction} \label{sec: introduction}
The interaction between a two-level system (qubit) and a harmonic oscillator (resonator) has been widely studied, originally as one of the simplest systems to study light-matter interaction~\cite{Rabi1937, Jaynes1963}, and later as a platform for quantum optics~\cite{Walls2008,Gleyzes2007,Goppl2008} and quantum information processing~\cite{Blais2004, Blais2007,Billangeon2015,Billangeon2015a}.
The deep strong coupling (DSC) regime of the qubit-resonator (Q-R) interaction, where the coupling strength $g$ is comparable to or even larger than the transition energies of the qubit ($\Delta$) and the resonator ($\omega_{\rm r}$), has recently been achieved using artificial atoms in superconducting circuit QED systems~\cite{Yoshihara2017,Yoshihara2017a,Yoshihara2018} and THz metamaterials coupled to the cyclotron resonance of a 2D electron gas~\cite{Bayer2017}, as reviewed in Refs.~\cite{FriskKockum2019,Forn-Diaz2019}.

In the DSC regime, the ground state of the Q-R system is quite different from that for weaker coupling~\cite{Ashhab2010,Rossatto2017}.
First, it is an entangled state between qubit and resonator.
Second, such a Schr\"odinger's cat-like state has a nonzero expectation value of the photon number, $\braket{n}=|g/\omega_{\rm r}|^2$.
These photons are referred to as virtual photons, since the system is in the ground state and therefore the photons cannot be spontaneously emitted.
Such nonclassical properties of the ground state are proposed to be useful for quantum metrology~\cite{Facon2016} and the preparation of nonclassical states of photons~\cite{Gheeraert2017,Leroux2017, Gu2017}.

Any quantum system realized in an actual experimental setup, however, is coupled to external degrees of freedom, intentionally or unintentionally.
Particularly, in a superconducting circuit QED system, transmission lines are usually attached to the resonators and the qubits for control and measurement.
A standard prescription for treating an open quantum system in the DSC regime is the Lindblad master equation in the dressed state picture~\cite{Beaudoin2011}, in which the system relaxes to the ground state of the system Hamiltonian at zero temperature.
However, the master equation is based on the Born-Markov approximation, which assume that the system-environment interaction is sufficiently weak so that the system and the environment are always in a product state and the state of the environment does not change during the time evolution.
When there is a nonzero interaction between the system and the environment, however, the total ground state can be entangled~\cite{Ashhab2006} and its reduced density matrix for the system is not necessarily the ground state of the system Hamiltonian.
It is not clear to what extent the Born-markov approximation is valid, or how robust the ground state properties of the DSC system are.
In particular, the energy gap between the ground state and the first excited state becomes exponentially small as $\Delta\expo{-2g^2/\omega_{\rm r}^2}$ with increasing  $g$~\cite{Nataf2010a}.
So the nonclassicality of the ground state is expected to be fragile against the coupling to an environment in the DSC regime, although the average number of virtual photons is shown to be only quantitatively affected by losses~\cite{DeLiberato2017}.
By increasing $g$, there is a competition between two effects: the increase of virtual photons, which enhances the nonclassicality, and the exponential decrease of the energy gap ($\sim\Delta\expo{-2g^2/\omega_{\rm r}^2}$), which degrades the nonclassicality due to the increase of fragility, in any realistic setting.
Because of this competition, it is not clear whether just increasing the coupling strength $g$ is helpful to obtain the maximum nonclassicality in the presence of an environment, even at zero temperature.

In this paper, we investigate the ground state properties of an open DSC system.
For this purpose, we propose a variational ground state for the enlarged qubit-resonator-environment system, which we call coherent variational state (CVS),  by extending the qubit-state-dependent coherent  state~[Eq.~\eref{cat state}] to the total system including the environmental degrees of freedom.
Based on the analysis with the  CVS and the numerical diagonalization of the truncated total Hamiltonian, we find that the effect of the coupling to the environment strongly depends on how the system is coupled to the environment, i.e., inductively or capacitively.
This strong dependence results from the fact that the ground state of the Q-R system in the DSC regime breaks rotational symmetry around the origin in the phase space.
When the resonator couples to the qubit and the environment in the same way (for instance, both inductively), we find that the average number of virtual photons increases, and that the quantum superposition realized in the Q-R system is partially degraded.
Furthermore, the ground state superposition tends to be more fragile when the Q-R coupling $g$ is larger, so that the nonclassicality of the resonator state, measured by the metrological power~\cite{Kwon2019}, is maximized at a moderate strength of the Q-R coupling.
When the resonator couples to the qubit inductively and to the environment capacitively, on the other hand, we did not observe any peak of the metrological power in the parameter region that we investigated, so that enhancement in metrological power is achievable.

This paper is organized as follows.
In Sec.~\ref{sec: model}, we describe the theoretical model and present examples of superconducting circuits realizing the Hamiltonian.
In Sec.~\ref{sec: CVS}, we outline the variational method based on the CVS.
In Sec.~\ref{sec: numeric}, by using the CVS and the numerical diagonalization, we numerically evaluate the number of virtual photons, the purity in the Q-R system, and the degree of nonclassicality measured by the metrological power.
In Sec.~\ref{sec: stability}, we discuss how the coupling-type dependence arises in the DSC system.
We conclude the paper in Sec.~\ref{sec: conclusion}.
In \ref{sec: circuit hamiltonian}, we describe the detailed derivation of the Hamiltonian from the circuit model.
In \ref{sec: spin-boson model}, we discuss the relation to the spin-boson model.
In \ref{sec: symmetry}, we discuss  how the symmetry of the total Hamiltonian restricts the form of the  CVS.
In \ref{sec: stationary}, we show that the stationary equations for the CVS can be substantially simplified by introducing a collective variable.
In \ref{sec: validity}, we check the validity of the CVS in the inductive coupling case by comparing the result from the numerical diagonalization.

\section{Model} \label{sec: model}
\begin{figure}[htbp]
\begin{center}
\includegraphics[width = 0.6\columnwidth]{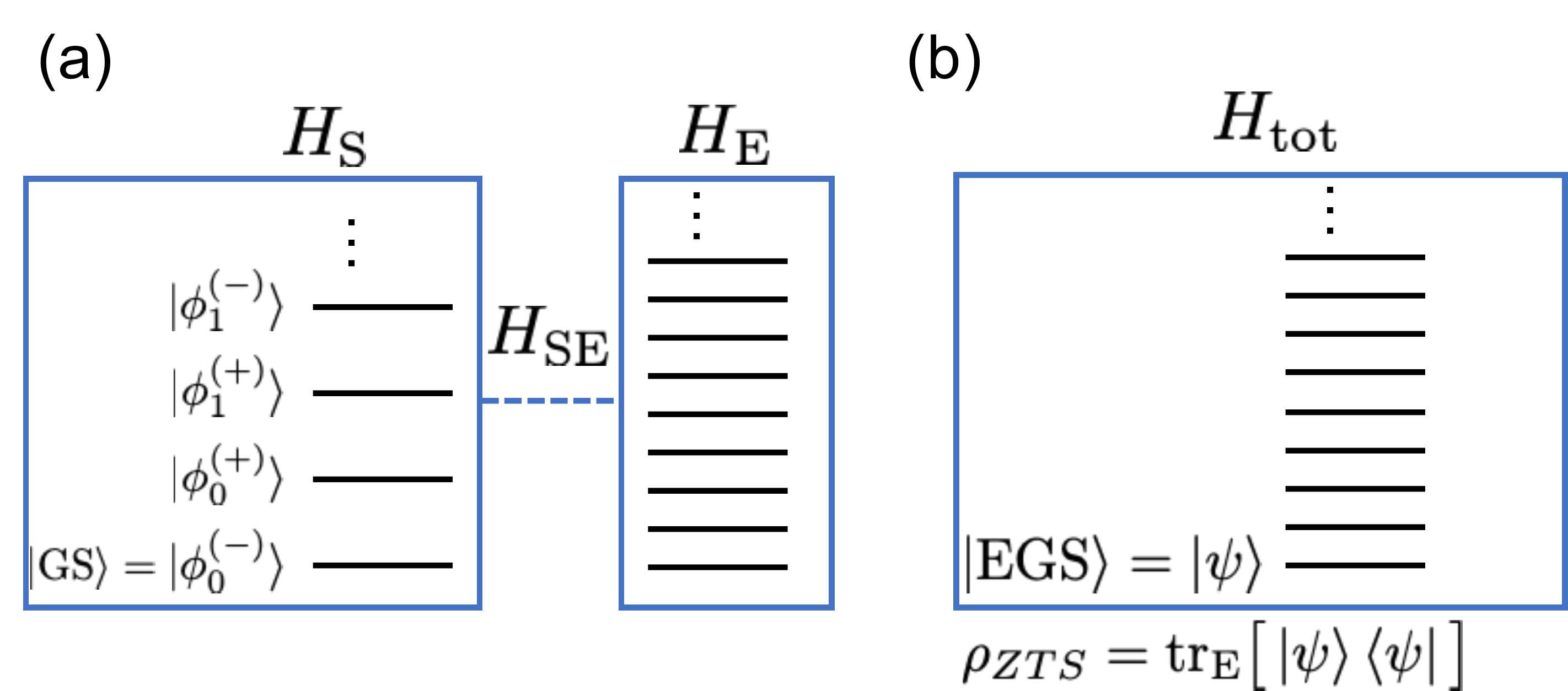}
\caption{Energy diagrams of the qubit-resonator-waveguide system. 
(a) The Q-R system (described by $H_{\rm S}$) for $g >\omega_{\rm r}/2$ and the environmental waveguide system (described by $H_{\rm E}$) are coupled with $H_{\rm SE}$.
(b) We denote the enlarged ground state of $H_{\rm tot}=H_{\rm S}+H_{\rm E}+H_{\rm SE}$ (qubit-resonator-waveguide) by $\ket{\rm EGS}=\ket\psi$. 
Its reduced matrix to the Q-R space is referred to as the zero-temperature state and is denoted by $\rho_{ZTS}={\rm tr}_{\rm E}\big[ \ket\psi\bra\psi \big]$.
\label{EnergyDiagram}}
\end{center}
\end{figure}

In this section, we describe the model considered in this paper.
The relationship among the system, the environment, and the total system is summarized in Fig.~\ref{EnergyDiagram}.

\subsection{Qubit and resonator}
We consider the quantum Rabi model, which is described by the Hamiltonian
\eq{
H_{\rm S}&=\omega_{\rm r}a^\dagger a + \frac{\Delta}{2}\sigma_x + g\sigma_z (a+a^\dagger). \label{system Hamiltonian}
}
Here, $a$ $(a^\dagger)$ is the annihilation (creation) operator of a resonator photon with energy $\omega_{\rm r}$, $\sigma_j$ ($j=x,y,z$) is the Pauli operator of the qubit with transition energy $\Delta(>0)$, and $g$ is the coupling strength between them.
The Planck constant $\hbar$ is set to unity throughout this paper.
Here,  we assume that  $X_I=a+a^\dagger$ is  proportional to the flux operator, so that the Q-R coupling is inductive.
When the coupling is capacitive, $X_I$ is replaced with $X_C=(a-a^\dagger)/i$ as
\eq{
H_{\rm S}'&=\omega_{\rm r}a^\dagger a + \frac{\Delta}{2}\sigma_x -i g\sigma_z (a-a^\dagger),
}
after choosing an appropriate basis for qubit states.
Both $H_{\rm S}$ and $H_{\rm S}'$ are unitary equivalent through a gauge transformation $a\rightarrow ia$, where the unitary operator is explicitly given by $U=\expo{\pm i\pi a^\dagger a/4}$.
We note that there is a subtlety in the coupling type (inductive/capacitive) because of the gauge ambiguity~\cite{DeBernardis2018a,DiStefano2019,Roth2019}.
Here,  the inductive/capacitive coupling refers to the Hamiltonian after choosing the optimal gauge with which the truncated Hamiltonian approximates the original gauge-invariant Hamiltonian well.

When $\omega_{\rm r} \gg \Delta$ or $g \gg \omega_{\rm r},\Delta$, the low-lying eigenstates of the quantum Rabi model are approximately described by~\cite{Ashhab2010, Rossatto2017}
\eq{
\ket{\phi_n^{(\pm)}(\alpha)} =  \frac{1}{\sqrt{2}}\left( \ket\uparrow\otimes D(-\alpha)\ket n \pm \ket\downarrow\otimes D(\alpha)\ket n \right), \label{approx eigenstate}
}
which is also denoted as $\ket{\phi_n^{(\pm)}}$.
Here, $\ket\uparrow$ ($\ket\downarrow$) is the qubit eigenstate corresponding to $\sigma_z=1$ ($-1$), $\alpha=g/\omega_{\rm r}$ is the amplitude of displacement in the resonator, and $D(\alpha)=e^{\alpha a^\dagger-\alpha^* a}$ is the displacement operator.
These parameter regions are referred to as the adiabatic oscillator limit~\cite{Ashhab2010, Irish2005,Irish2007,Albert2011} or the perturbative DSC regime~\cite{Forn-Diaz2019, Rossatto2017}, in the sense that the perturbation in terms of $\Delta$ is valid.
The eigenenergy is then, up to first order in $\Delta/\omega_{\rm r}$, given by
\eq{
E_n^{\pm}\simeq n\omega_{\rm r}-\frac{g^2}{\omega_{\rm r}}\pm\frac{\Delta}{2}\braket{n|D(2\alpha)|n} =n\omega_{\rm r}-\frac{g^2}{\omega_{\rm r}}\pm\frac{\Delta \ex^{-2\alpha^2}}{2}L_n(4\alpha^2),
}
where $L_n(x)$ is the Laguerre polynomial of the $n$-th order.
We note that the ground state, 
\eq{
\ket{{\rm GS}}=\ket{\phi_0^{(-)}}=\frac{1}{\sqrt{2}}\left( \ket\uparrow\otimes \ket{-\alpha} - \ket\downarrow\otimes\ket{\alpha} \right), \label{cat state}
}is the superposition of two coherent states displaced in opposite directions depending on the qubit state.

\subsection{Coupling to the environment}
\begin{figure}[htbp]
\begin{center}
\includegraphics[width = 0.8\columnwidth]{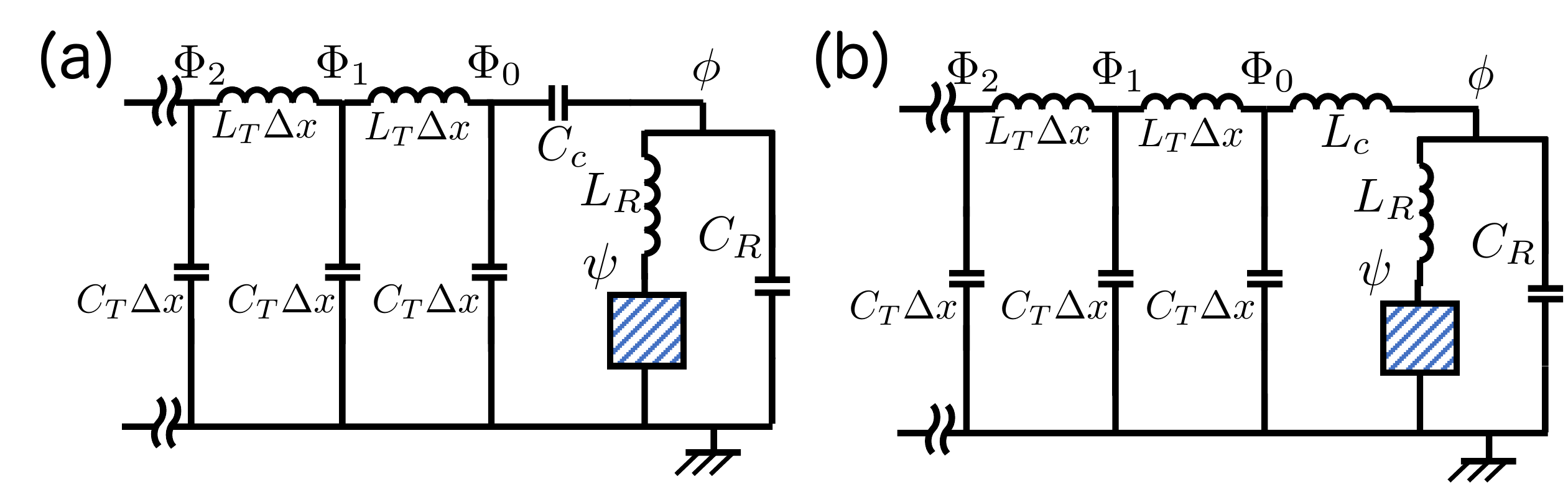}
\caption{Circuit diagrams of a Q-R system coupled to a waveguide. 
The blue shaded area is an arbitrary circuit representing the qubit.
The resonator-waveguide coupling is mediated by (a) a capacitance $C_c$ or (b) an inductance $L_c$.
\label{Circuit}}
\end{center}
\end{figure}

The environment can be modeled by an ensemble of harmonic oscillators~\cite{Leggett1987} as
\eq{
H_{\rm E}&=\sum_k\omega_kb_k^\dagger b_k,
}
where $b_k$ $(b_k^\dagger)$ is the annihilation (creation) operator of the $k$-th mode with energy $\omega_k$.
Each mode of the environment interacts with a system operator $X$ with strength $\xi_k$ as
\eq{
 H_{\rm SE}&=\sum_k\xi_kX(b_k+b_k^\dagger).  \label{interaction}
}
The total system is then described by the Hamiltonian $H_{\rm tot}=H_{\rm S} + H_{\rm E} + H_{\rm SE}$.

As a concrete model of an open DSC system, we investigate a DSC system coupled to a waveguide through the resonator (Fig.~\ref{Circuit}).
The shaded area can be an arbitrary circuit constituting a qubit, such as a Cooper pair box or a flux qubit.
The waveguide is coupled to the Q-R system through (a) a capacitance $C_c$ or (b) an inductance $L_c$.
When the interaction is mediated by an inductance (a capacitance), the system operator $X$ is the quadrature operator of the resonator, given as $X_I=a+a^\dagger$ ($X_C=(a-a^\dagger)/i$).
We note that in the case of the capacitive coupling, the interaction Hamiltonian is given by $H_{\rm SE}=-\sum_k\xi_k(a-a^\dagger)(b_k-b_k^\dagger)$. This is equivalent to Eq.~\eref{interaction} under the unitary transformation $b_k\rightarrow -ib_k$.
While the phase of the quadrature operator for $b_k$ does not affect the physical result because of the gauge invariance of $H_{\rm E}$, the phase of the quadrature operator for $a$ has the physical influence since it is used to couple the resonator to not only the waveguide but also the qubit.
In the following, we always assume that the Q-R coupling is inductive, and we analyze the cases where the resonator-waveguide (R-W) coupling is inductive or capacitive, except for Sec.~\ref{sec: stability}, where we discuss the relativity of the coupling.

Assuming a finite length $L$ of the waveguide, the wavenumber $k$ of the waveguide modes is discretized as $k=n\pi/L$ ($n\in\mathbb N$), and the energy is given by $\omega_k=vk$, where $v$ is the speed of microwave fields in the waveguide.
The coupling constant $\xi_k$ is given by
\eq{
\xi_k=\xi_0 \sqrt{\frac{\omega_k}{1+(\omega_k/\omega_{\rm cutoff})^2}\times \frac{\pi}{L}}\   \label{interaction spectrum}
}
in both the capacitive~\cite{Bamba2014} and inductive coupling case.
We do not have to put the cutoff factor by hand, since it is naturally included in the Hamiltonian derived from the circuit, and the cutoff energy $\omega_{\rm cutoff}$ is determined from the circuit parameters [Eqs.~\eref{cutoff_I} and~\eref{cutoff_C}].
In the small frequency region ($\omega_k \ll \omega_{\rm cutoff}$), the squared coupling strength $|\xi_k|^2$ is proportional to $\omega_k$, which corresponds to the Ohmic case in the spin-boson model (See~\ref{sec: spin-boson model} for details).

The loss rate $\kappa$ of a bare resonator photon into the waveguide is determined by the Fermi Golden rule as
\eq{
\frac{\kappa}{2\pi}= \xi_0^2\frac{\omega_{\rm r}}{1+(\omega_{\rm r}/\omega_{\rm cutoff})^2}. \label{loss rate formula}
}

\section{Coherent variational state} \label{sec: CVS}
In this section, we introduce the CVS and analyze the ground state of $H_{\rm tot}$.
In analogy to the approximate ground state of the quantum Rabi model [Eq.~\eref{cat state}],
we define the CVS of the total system as
\eq{
\ket{\psi_C(\alpha,\{\beta_k\})}=\frac{1}{\sqrt{2}}(\ket\uparrow\otimes\ket{-\alpha;\{-\beta_k\} } - \ket\downarrow\otimes\ket{\alpha;\{\beta_k\} }),  \label{Def:CoherentVariationalState}
}
where $\ket{\alpha;\{\beta_k\}}$ is the product of coherent states of the resonator and waveguide modes, satisfying $a\ket{\alpha;\{\beta_k\}}=\alpha\ket{\alpha;\{\beta_k\}}$ and $b_k\ket{\alpha;\{\beta_k\}}=\beta_k\ket{\alpha;\{\beta_k\}}$ for each $k$.
The variational parameters for the CVS are $\alpha$ and $\beta_k$.
We note that a more general form $c_0\ket\uparrow\otimes\ket{\alpha;\{\beta_j\} }+ c_1\ket\downarrow \otimes\ket{\alpha';\{\beta'_j\} }$ leads to the same results as the simpler form in Eq.~\eref{Def:CoherentVariationalState} due to the parity symmetry of the quantum Rabi Hamiltonian, as discussed in \ref{sec: symmetry}.
We also note that the renormalization of the qubit energy and the Rabi oscillation were analyzed using a similar ansatz by performing a polaron transformation ~\cite{Zueco2019}.

The total energy for the CVS is given by 
\eq{
E_{\rm CVS}&=& \braket{{\psi_C(\alpha,\{\beta_k\})}| H_{\rm tot}|{\psi_C(\alpha,\{\beta_k\})}}  \nonumber\\
&=&\omega_{\rm r}|\alpha|^2 - g(\alpha+\alpha^*) + \sum_k \omega_k |\beta_k|^2 \nonumber\\
&&\pm \sum_k\xi_k(\alpha\pm\alpha^*)(\beta_k\pm\beta_k^*)  - \frac{\Delta}{2}\expo{-2(|\alpha|^2 +\sum_k|\beta_k|^2)},
}
where the plus (minus) sign represents the inductive (capacitive) coupling.
The approximate ground state of the total system is the CVS $\ket{\psi_C(\bar{\alpha},\{\bar{\beta_k}\})}$, where $\bar{\alpha}$ and $\bar{\beta_k}$'s are the variational parameters that minimize the total energy $E_{\rm CVS}$.
Although there are a large number of degrees of freedom due to the numerous waveguide modes, the problem can be simplified into a stationary state problem with only two unknown parameters, $\alpha$ and $S=\sum_k|\beta_k|^2,$ as discussed in \ref{sec: stationary}.
Here, $S$ is a collective variable for the waveguide modes, representing the total number of virtual photons in the waveguide modes.

Once $\bar{\alpha}$ and $\bar{S}=\sum_k |\bar\beta_k|^2$ are obtained, the reduced density operator for the system, $\rho_{\scalebox{0.65}{ZTS}}={\rm tr}_{\rm E}[\ket{\psi_C(\bar{\alpha},\{\bar{\beta_k}\})}\bra{\psi_C(\bar{\alpha},\{\bar{\beta_k}\})}]$, which we call the zero-temperature state, is completely characterized by two parameters $\bar{\alpha}$ and $C=\expo{-2\bar{S}}$.
It is explicitly expressed as
\eq{
\rho_{\scalebox{0.65}{ZTS}}=&\frac{1+C}{2}\ket{\phi_0^{(-)}(\bar{\alpha})}\bra{\phi_0^{(-)}(\bar{\alpha})} + \frac{1-C}{2}\ket{\phi_0^{(+)}(\bar{\alpha})}\bra{\phi_0^{(+)}(\bar{\alpha})}.  \label{reduced density matrix}
}
This equation implies that the R-W coupling has two effects.
First, the displacement $\bar{\alpha}$ is modified, which means that the average number of virtual photons $|\bar{\alpha}|^2$ is changed.
In fact, as we will see in Sec.~\ref{sec: numeric}, the virtual photons increases as the R-W coupling increases.
Second, $\rho_{\scalebox{0.65}{ZTS}}$ includes the first excited state of $H_{\rm S}$ with a fraction $P_e=\frac{1-C}{2}$.
A similar behavior can be found in the case of a single two-level system coupled to an environment~\cite{Leggett1987}.

The quantity $C$ serves as a measure of coherence, which is a real quantity within the range of $0\le C\le 1$.
If $\beta_k=0$ and hence $C=1$, the system is in a pure state.
On the other hand, if $C=0$, the quantum superposition is completely destroyed and the reduced density matrix $\rho_{\rm S}$ is maximally mixed.
When we represent $\rho_{\rm S}$ in the basis of $\ket\uparrow\otimes\ket{-\bar{\alpha}}$ and $\ket\downarrow\otimes\ket{\bar{\alpha}}$, the quantity $C$ appears in the off-diagonal element:
\eq{
\rho_{\scalebox{0.65}{ZTS}}=\frac{1}{2}\left(
    \begin{array}{cc}
      1 & -C \\
      -C & 1 
    \end{array}
  \right).
}
In this sense, R-W coupling reduces the coherence realized in this basis.

As for the validity of the coherent variational state, we note that it gives the exact ground state of the total Hamiltonian if the qubit energy $\Delta$ and the counter-rotating terms $\xi_k( ab_k + a^\dagger b_k^\dagger )$ are neglected.
In Ref.~\cite{Yoshihara2017}, it is argued that the Q-R state $\ket{\phi_0^{(-)}}$ gives a rather accurate description of the ground state of the quantum Rabi Hamiltonian even when the qubit energy $\Delta$ is finite.
Furthermore, we compare the result of CVS in the inductive coupling case with that of the numerical diagonalization for a few waveguide mode case in \ref{sec: validity}, and show that the CVS describes not only the virtual photons but also the nonclassical properties well in the presence of the R-W coupling.

\section{Numerical calculations} \label{sec: numeric}
In this section, we numerically investigate the properties of $\rho_{\scalebox{0.65}{ZTS}}$ based on the CVS.
We adopt the bare resonator photon loss rate $\kappa$ [Eq.~\eref{loss rate formula}] as a measure of the R-W coupling strength.
The other parameters are set to be $\omega_{\rm r}/2\pi=6$ GHz and $\Delta/2\pi=1.2$ GHz.
See~\ref{sec: circuit hamiltonian} for details.
We again note that the Q-R coupling is assumed to be inductive in this section, and we refer to the coupling between the resonator and the waveguide when we mention the inductive or capacitive coupling.

\subsection{Inductive R-W coupling}

\begin{figure}[htbp]
\begin{center}
\includegraphics[width = 0.8\columnwidth]{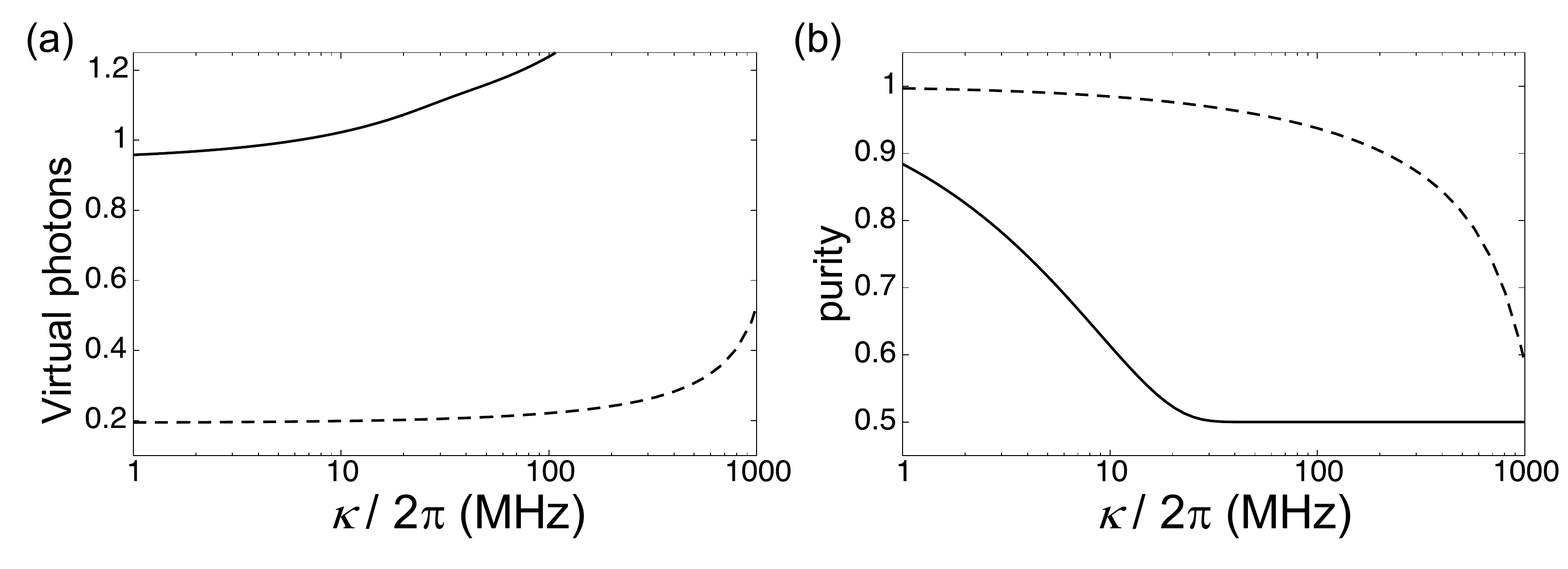}
\caption{The average virtual photon number (a) and the purity (b) plotted against the bare loss rate $\kappa$ in the inductive coupling case.
The Q-R coupling is $g/2\pi=$ 3 GHz (dashed line) and 6 GHz (solid line).
\label{Inductive_VP}}
\end{center}
\end{figure}

We first calculate the average number of virtual photons $|\bar{\alpha}|^2$ in the inductive coupling case.
Figure~\ref{Inductive_VP} (a) shows that, in the inductive coupling case, the number of virtual photons increase as the R-W coupling $\kappa$ increases.
Indeed, the interaction term $\sum_k\xi_k(a+a^\dagger)(b_k+b_k^\dagger)$ acts as a shifter on the resonator phase space in the real direction.
Furthermore, we can show that the energy gradient $ \partial E/\partial\alpha$ takes a negative value at $\alpha=\tilde\alpha$, where $\tilde\alpha$ is the stationary solution in the absence of the R-W coupling, which implies that the average number of virtual photons always increases due to the R-W coupling, independent of the details of the interaction spectrum $\xi_k$.

Next, we calculate the purity of $\rho_{\scalebox{0.65}{ZTS}}$, defined by
\eq{
\gamma=\tr[\rho_{\scalebox{0.65}{ZTS}}^2].
}
Figure~\ref{Inductive_VP} (b) shows that, in the inductive coupling case, the purity decreases as the loss rate $\kappa$ increases.
By comparing the results for $g/2\pi=$ 3 and 6 GHz, we see that the purity also decreases as the Q-R coupling $g$ increases.
In other words, in the DSC regime, the quantum coherence of the ground state becomes fragile when the Q-R interaction $g$ is extremely large.
It implies that, even though the exact ground state of the quantum Rabi model becomes increasingly useful with increasing $g$, for instance, for quantum metrological tasks~\cite{Facon2016}, the maximum performance is achieved at a moderate strength of the interaction $g$ when the coupling to the environment is taken into account.
Indeed, the nonclassicality, measured by the metrological power~\cite{Kwon2019}, has the maximum at a certain value of $g$, as we discuss below.

\begin{figure}[htbp]
\begin{center}
\includegraphics[width = 0.8\columnwidth]{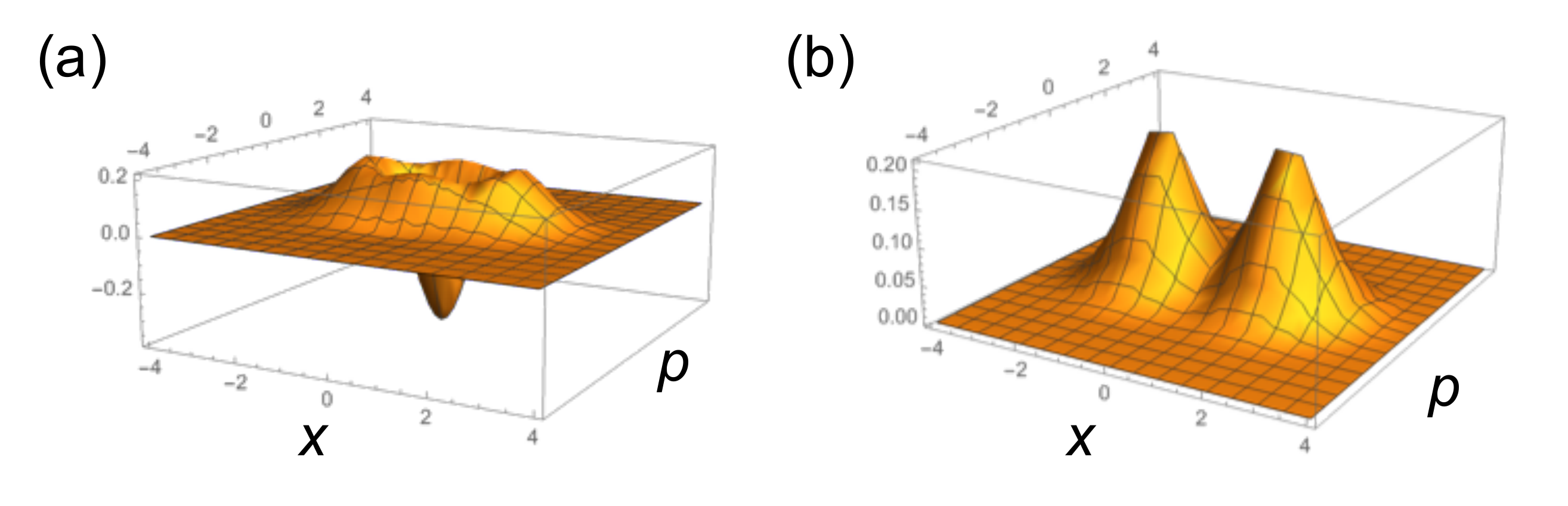}
\caption{The Wigner functions of the post-measurement state $\rho_{\sigma_x=-1}$ in the inductive coupling case for $g/2\pi=6$ GHz. (a) $\kappa/2\pi=1$ MHz. The state possesses sufficient coherence ($C=0.88$) for the Wigner function to take a negative value around the origin in phase space. (b) $\kappa/2\pi= 40$ MHz. The state is almost decohered ($|C|<10^{-4}$), and the Wigner function is that of the mixture of two coherent states.
\label{WignerFunc}}
\end{center}
\end{figure}
Let us evaluate the ``quantumness" of  the ground state of the system.
There are many ways of defining and quantifying the quantumness~\cite{Kwon2019,Genoni2010,Takagi2018,Albarelli2018,Yadin2018}.
Here, we project the state onto the eigenstates of some qubit operator, and calculate the quantumness from the reduced density matrix of the resonator, by exploiting the resource theory of the nonclassicality in continuous variable systems.
One of the measures of the nonclassicality is the metrological power~\cite{Kwon2019}, which quantifies the maximum achievable quantum enhancement in displacement metrology based on the quantum Fisher information~\cite{Helstrom1968,Holevo2011}.
For a resonator state with the spectral decomposition $\rho_{\rm R}=\sum_i\lambda_i\ket i\bra i$, the elements of quantum Fisher information matrix for quadrature operators are defined as
\eq{
F_{kl}=2\sum_{i,j}\frac{(\lambda_i-\lambda_j)^2}{\lambda_i+\lambda_j}\braket{i|R^{(k)}|j}\braket{j|R^{(l)}|i} , \ (k,l=1,2),
}
where $R^{(1)}=(a+a^\dagger)/\sqrt{2}$ and $R^{(2)}=(a-a^\dagger)/\sqrt{2}i$ are the quadrature operators.
Then the metrological power of the resonator state $\rho_{\rm R}$ is given by
\eq{
{\mathcal M}(\rho_{\rm R})=\max\left\{ \frac{\lambda_{\rm max}(F)-1}{2} ,0 \right\},
}
where $\lambda_{\rm max}(F)$ is the maximum eigenvalue of the quantum Fisher information matrix.

We consider a projective measurement of a qubit operator
\eq{
\sigma_{\theta,\phi}=\sigma_x\sin\theta\cos\phi+\sigma_y\sin\theta\sin\phi+\sigma_z\cos\theta.
}
The measurement outcome $\sigma_{\theta,\phi}=\pm 1$ is obtained with probability 
\eq{
{\rm Prob}[\sigma_{\theta,\phi}=\pm1]= {\rm tr}\left[ P_{\sigma_{\theta,\phi}=\pm 1} \rho_{\scalebox{0.65}{ZTS}} \right],
}
and the post-measurement state for the resonator is
\eq{
\rho_{\sigma_{\theta,\phi}=\pm 1}\propto {\rm tr}_{\rm qubit}\left[ P_{\sigma_{\theta,\phi}=\pm 1} \rho_{\scalebox{0.65}{ZTS}} \right],
}
where $P_{\sigma_{\theta,\phi}=\pm 1} $ is the projection operator.
Physically, the post-measurement resonator state is a partially decohered cat state possessing interference fringes around the origin in the phase space representation.
These fringes become clearer as $\alpha$ and $|C|$ increase, which enables us to measure the displacement more precisely than coherent states (Fig.~\ref{WignerFunc}).
We can define the average metrological power as
\eq{
{\mathcal M}^{\rm av}(\rho_{\scalebox{0.65}{ZTS}}, \sigma_{\theta,\phi})=\sum_{a=\pm1} {\rm Prob}[\sigma_{\theta,\phi}=a]{\mathcal M}(\rho_{\sigma_{\theta,\phi}=a}).
}
Finally we define the metrological power of the Q-R system state by optimizing the measurement axis of the qubit as
\eq{
{\mathcal M}(\rho_{\scalebox{0.65}{ZTS}})=\max_{\theta,\phi} {\mathcal M}^{\rm av}(\rho_{\scalebox{0.65}{ZTS}}, \sigma_{\theta,\phi}). \label{MP for system}
}

\begin{figure}[htbp]
\begin{center}
\includegraphics[width = 0.8\columnwidth]{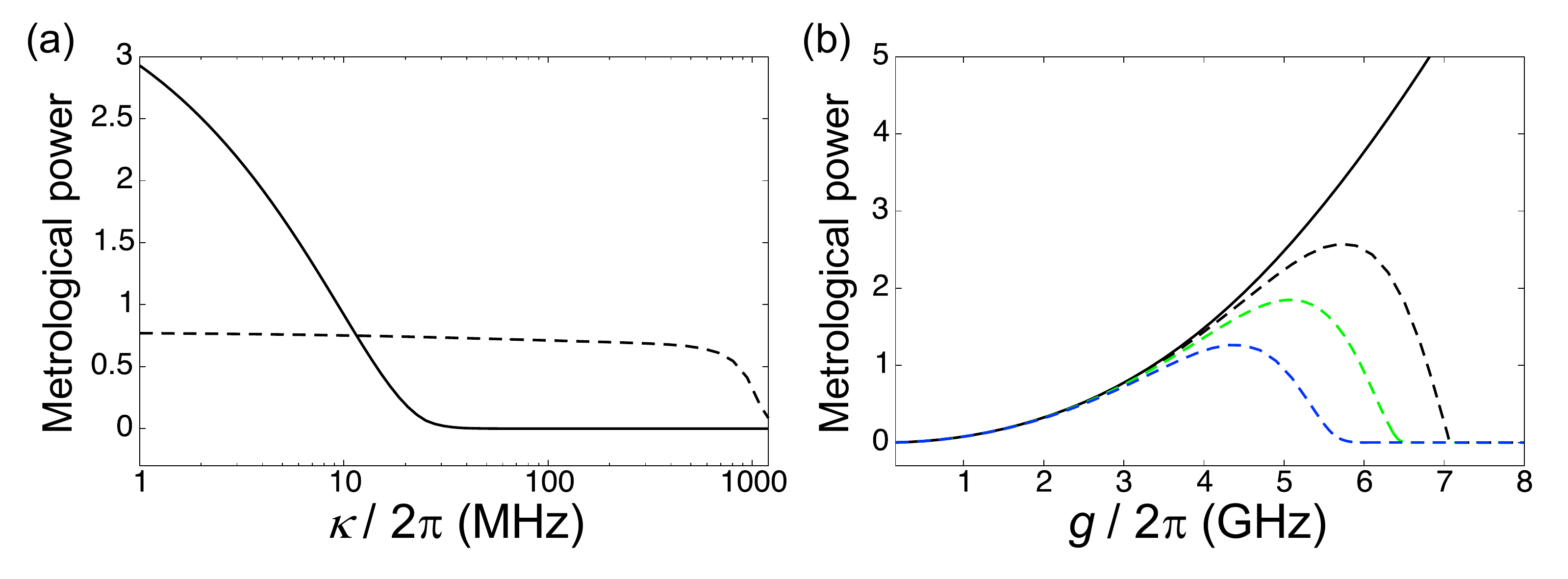}
\caption{
Metrological power ${\mathcal M}$ [Eq.~\eref{MP for system}] plotted against (a) the loss rate $\kappa$  and (b) the Q-R coupling $g$ in the inductive coupling case.
(a) The Q-R coupling is $g/2\pi=$ 3 GHz (dashed line) and 6 GHz (solid line).
(b) The bare loss rate is $\kappa/2\pi=$ 0 (black solid line), 2 MHz (black dashed line), 10 MHz (green dashed line), and 50 MHz (blue dashed line).
\label{Inductive_MP}}
\end{center}
\end{figure}

In Fig.~\ref{Inductive_MP}, the metrological power of $\rho_{\scalebox{0.65}{ZTS}}$~[Eq.~\eref{MP for system}] is plotted against (a) the loss rate $\kappa$ and (b) the Q-R coupling $g$.
In our setting, the average metrological power is found to be maximized at $\theta=\phi=\pi/2$, corresponding to the measurement of $\sigma_y$.
Figure~\ref{Inductive_MP} (a) shows that for each value of $g$, the metrological power rapidly decreases to zero when $\kappa$ becomes larger than a certain value.
This critical value of $\kappa$ becomes small as the Q-R coupling $g$ increases.
Figure~\ref{Inductive_MP} (b) shows that the average metrological power has a maximum at some finite value of $g$.
This maximum is achieved when the loss rate $\kappa$ is comparable to the energy gap $\Delta\ex^{-2g^2/\omega_{\rm r}^2}$, or $g_{\rm opt} \sim\omega_{\rm r}\sqrt{\log(\Delta/\kappa)/2}$, so that the optimal coupling strength $g_{\rm opt}$ increases only logarithmically by decreasing $\kappa$ and increasing $\Delta$.
In practice, the loss rate $\kappa$ cannot be too small because the measurement and control need time duration $T\sim1/\kappa$, during which decoherence occurs.
Therefore, this result implies that it is important to design a circuit to have a proper strength of the Q-R coupling $g$ and the loss rate $\kappa$ to obtain an optimal metrological advantage.

\subsection{Capacitive R-W coupling}
The capacitive coupling affects the system much less than the inductive coupling does, as we see below.
Indeed, in the capacitive coupling case, the CVS cannot capture the effect of R-W coupling, since it gives the exactly same result as the noninteracting case regardless of the coupling strength, as proved in~\ref{sec: stationary}.
Therefore, we compare the result from the CVS and the numerical diagonalization in the capacitive coupling case, and the CVS in the inductive coupling case.
To perform the numerical calculation, the total Hamiltonian is truncated as follows.
We take 14 photons and 3 photons into account for the resonator mode and each waveguide mode, respectively.
As the waveguide modes, we consider 4 modes with energies $\omega_k/2\pi=5,10,15,20$ GHz.
Although this truncation is not sufficient to quantitatively discuss the effect of the coupling to the environment, it shows the difference between the inductive coupling and the capacitive coupling.

\begin{figure}[htbp]
\begin{center}
\includegraphics[width = 0.8\columnwidth]{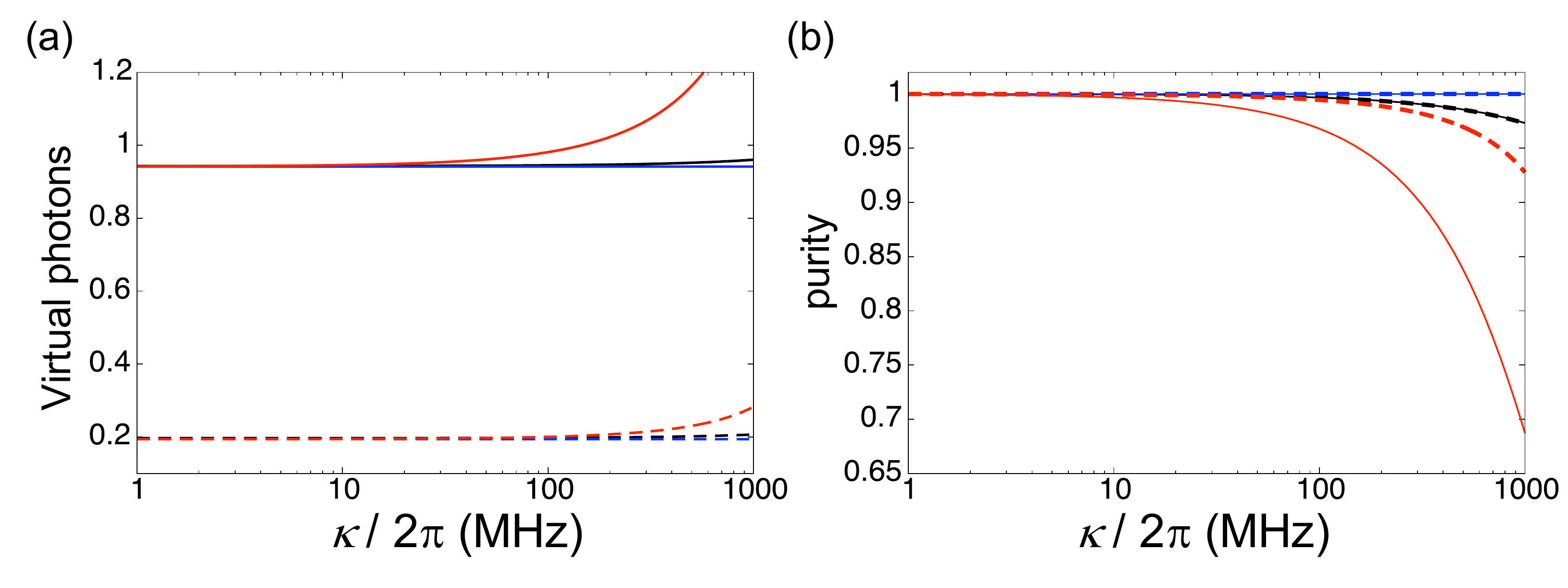}
\caption{The average virtual photon number (a) and the purity (b) plotted against the bare loss rate $\kappa$, calculated from the numerical diagonalization (black) and the CVS (blue) in the capacitive coupling case, and from the CVS in the inductive coupling case (red).
The Q-R coupling is $g/2\pi=$ 3 GHz (dashed line) and 6 GHz (solid line).
The results of the numerical diagonalization for the inductive coupling case are omitted here, and are shown in~\ref{sec: validity}.
\label{Capacitive_VP}}
\end{center}
\end{figure}

In Fig.~\ref{Capacitive_VP}, the average virtual photon number (a) and the purity (b) are plotted against the loss rate $\kappa$.
In the capacitive coupling case, the average number of virtual photons is much less sensitive to the R-W coupling compared to the inductive coupling case.
This fact can be qualitatively understood as follows.
The coupling operator $X_C=(a-a^\dagger)/i$ acts as a shifter of the displacement $\alpha$ in the imaginary direction.
However, since $\alpha=g/\omega_{\rm r}$ is real without the environment, the amplitude $|\alpha|^2$ is much less sensitive to the imaginary shift than the real shift.
We also see that in Fig.~\ref{Capacitive_VP} (b), the purity is less affected by the capacitive coupling to the waveguide compared to the inductive coupling case.
We will discuss the origin of this stability against the capacitive coupling to the waveguide in Sec.~\ref{sec: stability}.

\begin{figure}[htbp]
\begin{center}
\includegraphics[width = 0.8\columnwidth]{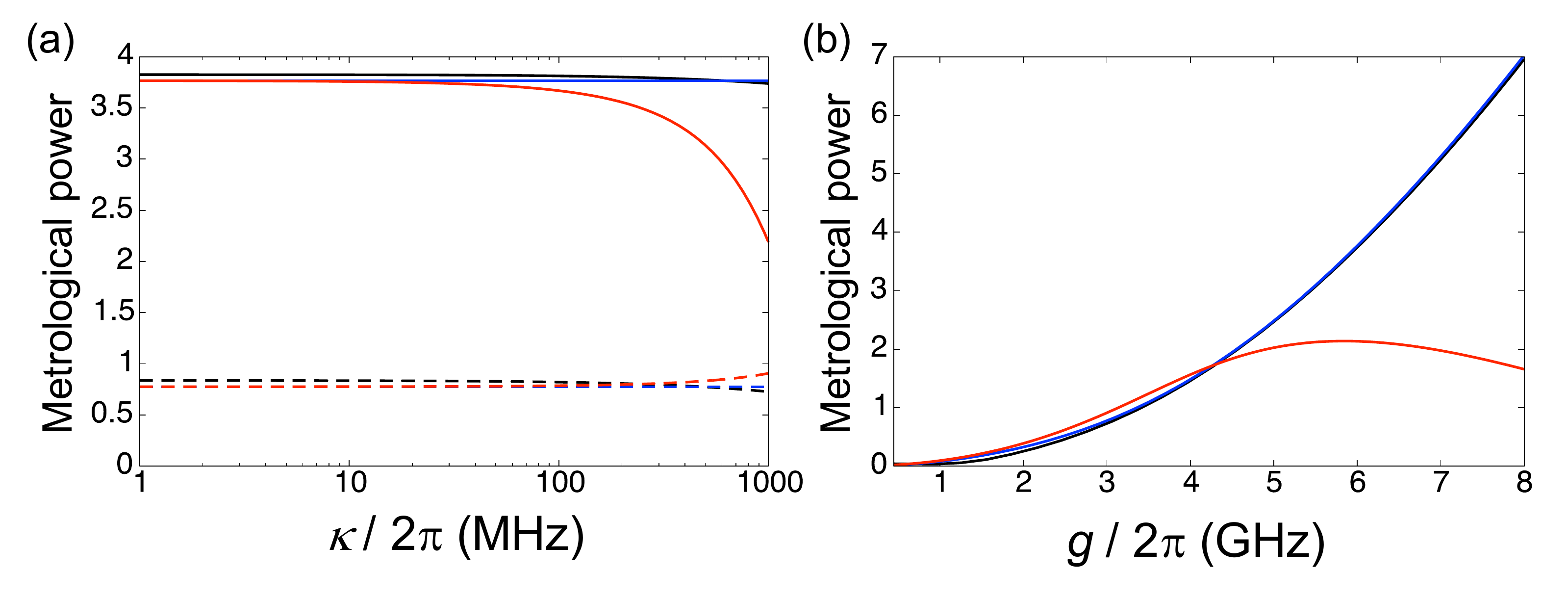}
\caption{Metrological power ${\mathcal M}$ [Eq.~\eref{MP for system}] plotted  against (a) the bare loss rate $\kappa$ and (b) the Q-R coupling $g$ calculated from the numerical diagonalization (black) and the CVS (blue) in the capacitive coupling case, and from the CVS in the inductive coupling case (red).
(a) The Q-R coupling is $g/2\pi=$ 3 GHz (dashed line) and 6 GHz (solid line).
(b) The loss rate is $\kappa/2\pi=$ 1000 MHz.
\label{Capacitive_MP}}
\end{center}
\end{figure}

In Fig.~\ref{Capacitive_MP}, the average metrological power is plotted against (a) the bare loss rate $\kappa$ and (b) the Q-R coupling $g$.
Since the effect of the W-R coupling is much underevaluated due to the truncation and the small number of environmental modes, we choose a rather huge value of $\kappa/2\pi=1000$ MHz, which is formally obtained from Eq.~\eref{loss rate formula}.
We see that in the capacitive coupling case, the metrological power calculated from the CVS and the numerical diagonalization agrees very well, and also that the nonclassicality is hardly affected by the capacitive coupling to the waveguide, compared to the inductive coupling case.

\section{Stability in the R-W capacitive coupling case}  \label{sec: stability}
In this section, we discuss the origin of the stability of the ground state when the R-W coupling is capacitive.
To obtain some physical insight, let us first consider the case where a bare resonator is coupled to a waveguide.
The fraction of the first excited state $\ket 1$ contained in the total ground state is proportional to $|\braket{1|X|0}|^2$ at the lowest order, where $X$ is a quadrature operator.
This transition amplitude is completely insensitive to the type of coupling, i.e., inductive $X_I=a+a^\dagger$ or capacitive $X_C=(a-a^\dagger)/i$, as 
\eq{
|\braket{1|X_I|0}|^2=|\braket{1|X_C|0}|^2=1.
}
This is due to the fact that the resonator Hamiltonian $H=\omega_{\rm r}a^\dagger a$, and the energy eigenstates $\ket n$ are invariant under the rotation around the origin in the phase space, represented by the unitary transformation $U=\exp\kakko{i\theta a^\dagger a}$.
On the other hand, since the eigenstates of a DSC system are not invariant under this rotation, the transition amplitude strongly depends on the type of coupling as 
\eq{
|\braket{\phi_0^{(+)}(\alpha)|X_I|\phi_0^{(-)}(\alpha)}|^2&=4|{\rm Re}[\alpha]|^2=4\kakko{\frac{g}{\omega_{\rm r}}}^2, \\
|\braket{\phi_0^{(+)}(\alpha)|X_C|\phi_0^{(-)}(\alpha)}|^2&=4|{\rm Im}[\alpha]|^2=0.  \label{matrix element capacitive}
}
Therefore, when the Q-R coupling is mediated by the inductance, i.e., $\alpha$ is real, the system is stable against capacitive coupling to the waveguide.
A similar argument is applied in Ref.~\cite{Nataf2011} to protect a qubit from noise based on the fact that the transition amplitude of the qubit operator becomes exponentially small in $g$.
In contrast, in our case, the transition amplitude is exactly zero when the R-W coupling is capacitive.

\begin{figure}[htbp]
\begin{center}
\includegraphics[width =0.9\columnwidth]{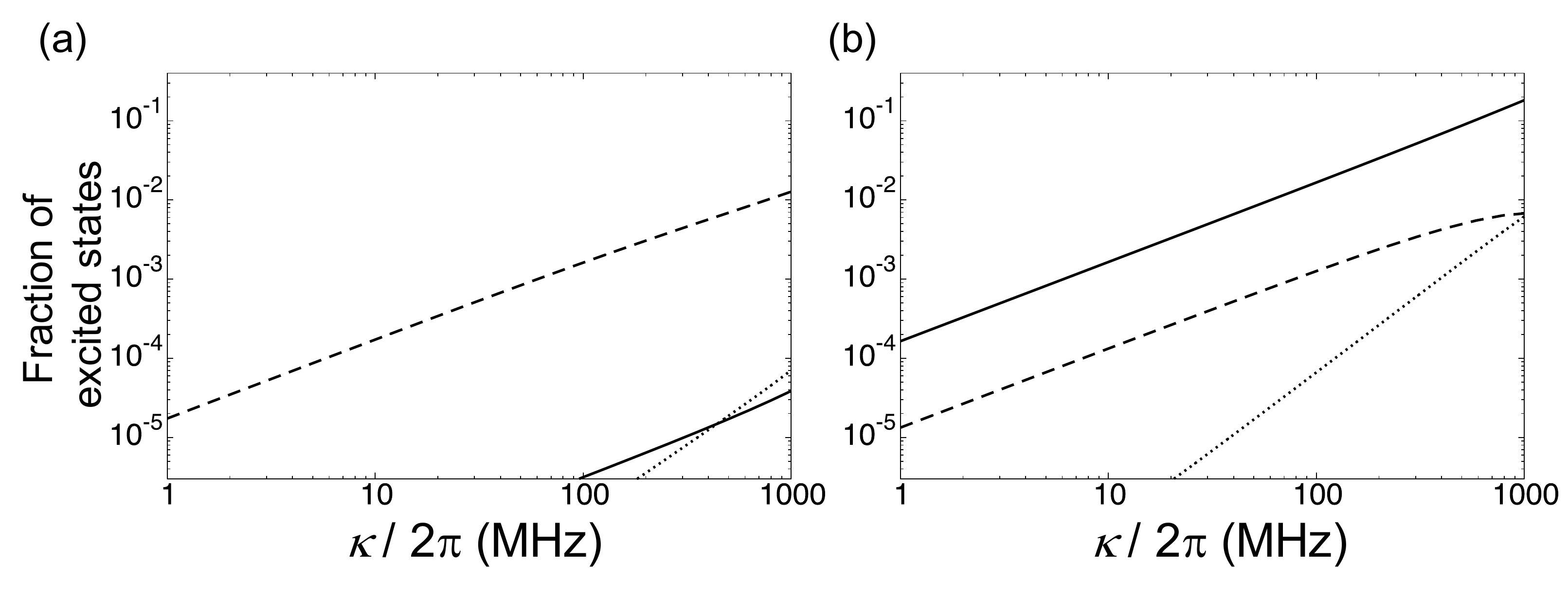}
\caption{The fraction of the excited states of the isolated system [Eq.~\eref{system Hamiltonian}] ($\ket{\phi_0^{(+)}}$ (solid line), $\ket{\phi_1^{(-)}}$ (dashed line) and $\ket{\phi_1^{(+)}}$ (dotted line)) contained in $\rho_{\scalebox{0.65}{ZTS}}$, calculated from the numerical diagonalization, plotted against the loss rate $\kappa$.
The R-W coupling is (a) capacitive or (b) inductive.
\label{circuit inductive}}
\end{center}
\end{figure}

To see this coupling-type-dependence more directly, we perform a numerical diagonalization of the truncated total Hamiltonian of the qubit-resonator-waveguide system.
The truncation is the same as in the previous section.
Figure~\ref{circuit inductive} shows the fraction of the excited states of the Q-R system contained in the ground state of $H_{\rm tot}$.
In the inductive coupling case, the most dominant excitation is the first excited state $\ket{\phi_0^{(+)}}$.
On the other hands, the fraction of $\ket{\phi_0^{(+)}}$ is not dominant in the capacitive coupling case, as is expected from Eq.~\eref{matrix element capacitive}.
Instead, the most dominant excited state is $\ket{\phi_1^{(-)}}$, which is the only excited state with a nonzero  transition amplitude as $|\braket{\phi_1^{(-)}(\alpha)|X_C|\phi_0^{(-)}(\alpha)}|^2=1$.

The reason why the system is not changed in the capacitive coupling case in the CVS analysis in Sec.~\ref{sec: numeric} is that the CVS considers only the two lowest eigenstates $\ket{\phi_0^{(-)}}$ and $\ket{\phi_0^{(+)}}$.
From the analysis performed in~Sec.\ref{sec: numeric}, we cannot determine whether the metrological power monotonically increases as a function of $g$ or peaks at a certain value of $g$ in the capacitive coupling case.
However, the peak, if it exists, is expected to occur at a much larger value of $g$ than in the inductive coupling case, so that a higher metrological power is achievable in the capacitive coupling case.

We need to be careful to conclude from our results that the capacitive R-W coupling is superior to the inductive coupling, since there is a tradeoff between the gate speed and the relaxation time~\cite{Koshino2020}.
In the capacitive coupling case, this tradeoff relation indicates that the long relaxation time implies a slow control between the ground state and the first excited state.

Finally, we stress that  only the relative phase between the resonator quadrature operators coupled to the qubit and the waveguide is relevant to this stability.
When the Q-R coupling is assumed to be capacitive, the system is sensitive to the capacitive coupling to the waveguide and insensitive to the inductive coupling.
These results are summarized in Table~\ref{Correspondence}.

\begin{table}[htbp]
\begin{center}
\caption{The transition amplitude between the ground state and the first excited state for several system Hamiltonians and different types of the R-W coupling. Pairs of columns connected by arrows are unitary equivalent.
\label{Correspondence}}
\qquad \includegraphics[width = 0.8\columnwidth]{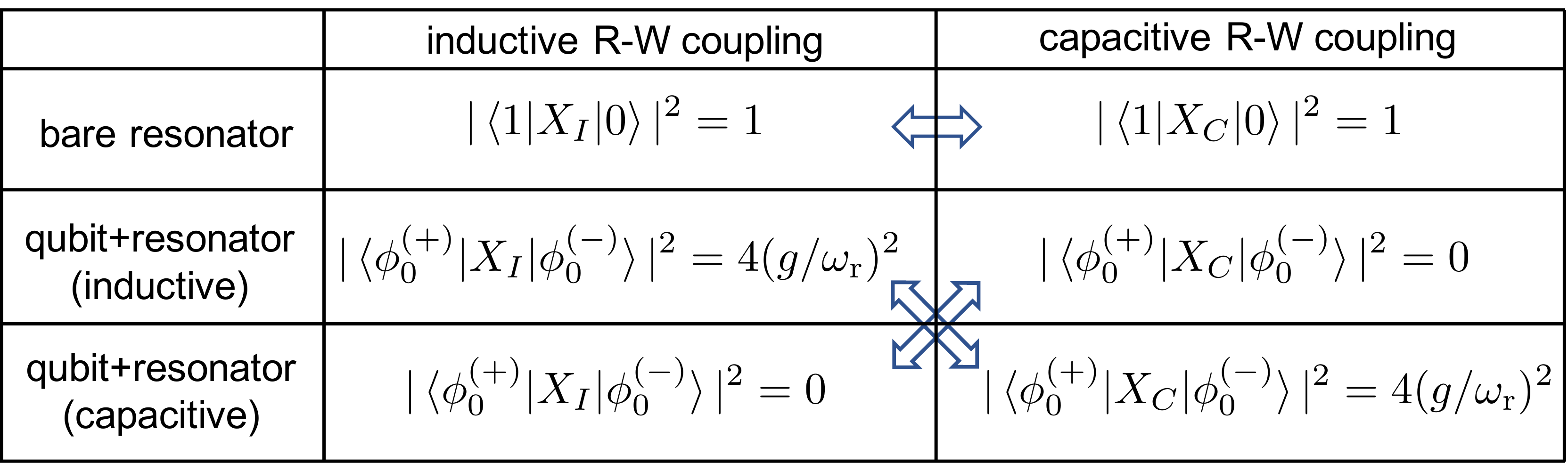}
\end{center}
\end{table}

\section{Conclusion} \label{sec: conclusion}
In this paper, by analyzing the  ground state of a qubit-resonator-waveguide system, we have investigated the effect of an environment on the ground state of the quantum Rabi model in the DSC regime.
We have introduced the qubit-state-dependent coherent variational state (CVS)~[Eq.~\eref{Def:CoherentVariationalState}].
This variational ansatz is easy to analyze and is consistent with the result from  numerical diagonalization.
We have shown that the zero-temperature state $\rho_{\scalebox{0.65}{ZTS}}$ strongly depends on the type of the R-W coupling because of the broken rotational symmetry in the eigenstates of the DSC system.

When the resonator couples to the qubit and the waveguide in the same way (for instance, both are inductive), the number of virtual photons increases due to the R-W coupling, which might be advantageous to detect virtual photons experimentally~\cite{Lolli2015,Munoz2018}.
We have also shown that, even at zero temperature, the Q-R system is a mixed state and contains the excited states of the quantum Rabi Hamiltonian, which implies the fragility of the quantum superposition realized in the ground state.
As a result, the nonclassicality of the resonator system, measured by the metrological power, is maximized at a certain coupling strength $g$, when the environment is taken into account.
We  note that the analysis based on the multi-polaron expansion~\cite{Bera2014} suggests that the CVS underestimate the coherence in the system.
To obtain a more accurate result in the inductive coupling case, we may modify the CVS to include more than one polaron.

On the other hand, when the resonator couples to the qubit and the waveguide in different ways (for instance, one is inductive and the other is capacitive), the system is almost unaffected, so that a higher metrological power than the same coupling case is achievable in the presence of environment.
It is worth considering a better variational ansatz that can quantitatively describe the ground state in such case.

Our results offer guiding principles to obtain a better metrological advantage when we design superconducting circuit QED systems. 
Since it is necessary to perform projective measurements on the qubit to exploit this metrological advantage, our results also demonstrate the advantages of achieving dynamically controllable coupling between qubit and resonator.


\ack
We would like to thank S. Masuda, R. Takagi, and I. Iakoupov for fruitful discussions.
This work was supported by Japan Science and Technology Agency (JST) Core Research for Evolutionary Science and Technology (CREST) Grant Number \mbox{JPMJCR1775} and JST Precursory Research for Embryonic Science and Technology (PRESTO) Grant number \mbox{JPMJPR1767}, Japan.

\appendix
\section{Derivation of the Hamiltonian of the circuit coupled to two waveguides}\label{sec: circuit hamiltonian}
\begin{figure}[htbp]
\begin{center}
\includegraphics[width =0.7 \columnwidth]{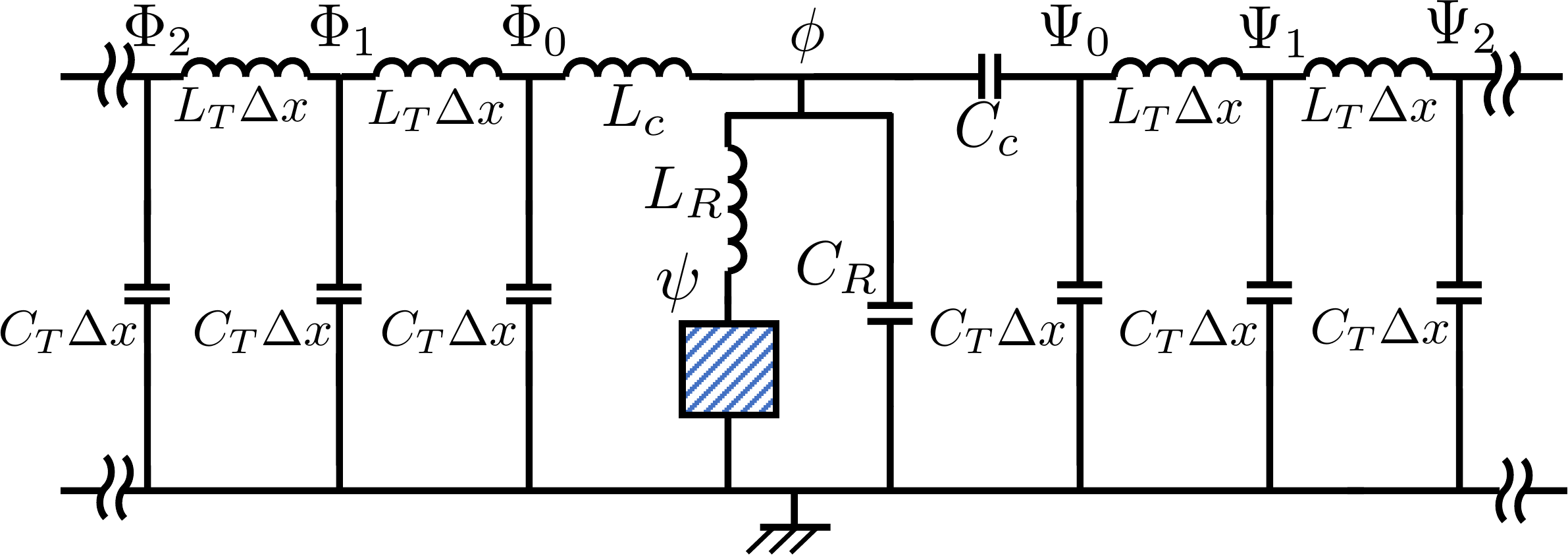}
\caption{Circuit diagram of the Q-R system coupled to two waveguides, one inductively (left) and one capacitively (right).
\label{both circuit}}
\end{center}
\end{figure}

In this section, we derive the Hamiltonians of the circuits in Fig~\ref{Circuit} (a) and (b).
For that purpose, we consider a Q-R circuit coupled to two waveguides, one inductively and one capacitively, as shown in Fig.~\ref{both circuit}.
The circuits in Fig~\ref{Circuit} (a) and (b) can be obtained by taking the limits $L_c\rightarrow\infty$ and $C_c\rightarrow0$, respectively.
We note that a similar circuit is discussed in Ref.~\cite{Parra-Rodriguez2018}.

We assign a flux variable to each vertex.
Here, variables $\psi$, $\phi$, $\Phi_j$ and $\Psi_j$ represent degrees of freedom of the qubit, the resonator, the waveguide coupled inductively (left), and the waveguide coupled capacitively (right), respectively.
The shaded area in Fig.~\ref{both circuit} can be an arbitrary circuit constituting a qubit, such as a Cooper pair box or a flux qubit.
We do not specify the details of the qubit circuit, but assume that it involves only $\psi$ and $\dot\psi$.
Then the total Lagrangian is given by
\eq{
&\mathcal L (\psi,\phi,\Phi_j,\Psi_j,\dot\psi,\dot\phi,\dot\Phi_j,\dot\Psi_j)\nonumber\\
=& \mathcal L_M (\psi,\dot\psi) - \frac{(\phi-\psi)^2}{2L_R}+\frac{C_R\dot\phi^2}{2}  \nonumber\\
&- \frac{(\phi-\Phi_0)^2}{2L_c}+\sum_{j=0}^\infty\kakko{\frac{C_T\Delta x\dot\Phi_j^2}{2} - \frac{(\Phi_{j+1}-\Phi_j)^2}{2L_T\Delta x}}\nonumber\\
&+ \frac{C_c }{2}(\dot\phi-\dot\Psi_0)^2+\sum_{j=0}^\infty\kakko{\frac{C_T\Delta x\dot\Psi_j^2}{2} - \frac{(\Psi_{j+1}-\Psi_j)^2}{2L_T\Delta x}}.
}

Canonical conjugate variables are defined as
\eq{
q&=\pd{1}{\mathcal L}{\dot\phi}=C_R\dot\phi  + C_c(\dot\phi-\dot\Psi_0) ,\\
\bar q&=\pd{1}{\mathcal L}{\dot\psi}=\pd{1}{\mathcal L_M}{\dot\psi},\\
Q_j&=\pd{1}{\mathcal L}{\dot\Phi_j}=C_T\Delta x\dot\Phi_j, \\
\bar Q_0&=\pd{1}{\mathcal L}{\dot\Psi_0}=C_c(\dot\Phi_0 - \dot\phi),\\
\bar Q_j&=\pd{1}{\mathcal L}{\dot\Psi_j}=C_T\Delta x\dot\Psi_j \ (j\ge1).
}
By performing the Legendre transformation, we obtain the Hamiltonian of the total circuit as
\eq{
H=&\kakko{\bar q\dot\psi-\mathcal L_M-\frac{\phi\psi}{L_R} + \frac{\psi^2}{2L_R} }
+\kakko{\frac{q^2}{2C_R} + \frac{\phi^2}{2L'_R} }\nonumber\\
&+\left[ \frac{\Phi_0^2}{2L_C} +\sum_{j=0}^\infty\kakko{\frac{Q_j^2}{2C_T\Delta x} + \frac{(\Phi_{j+1}-\Phi_j)^2}{2L_T\Delta x} } \right] \nonumber\\
&+\left[  \frac{\bar Q_0^2}{2C'_c} +\sum_{j=0}^\infty\kakko{\frac{\bar Q_j^2}{2C_T\Delta x} + \frac{(\Psi_{j+1}-\Psi_j)^2}{2L_T\Delta x} } \right] \nonumber\\
&-\frac{\phi\Phi_0}{L_c}+\frac{q\bar Q_0}{C_R}
}
where $L'_R=\frac{L_cL_R}{L_c+L_R}$ and $C'_c=\frac{C_cC_R}{C_c+C_R}$.
The first line represents the Hamiltonian for the qubit, the resonator, and the interaction between them.
The second (third) line represents the Hamiltonian for the waveguide inductively (capacitively) coupled to the resonator.
The final line represents the interaction between the resonator and the waveguide modes.
We note that the qubit variables $\{\psi,\bar q\}$ do not appear except for the first term line.
To diagonalize the waveguide modes, let us consider the equations of motion for the waveguide variables:
\eq{
\dot Q_j&=\frac{\Phi_{j+1} + \Phi_{j-1} -2\Phi_j }{ L_T\Delta x } \ (j\ge1),\\
\dot\Phi_j&=\frac{Q_j}{C_T\Delta x}  \ (j\ge1),\\
\dot Q_0&=-\frac{\Phi_0}{L_c}-\frac{\Phi_0-\Phi_1}{L_T\Delta x},\\
\dot\Phi_0&=\frac{Q_0}{C_T\Delta x},\\
\dot{\bar Q}_j&=\frac{\Psi_{j+1} + \Psi_{j-1} -2\Psi_j }{ L_T\Delta x } \ (j\ge1) , \\
\dot\Psi_j&=\frac{\bar Q_j}{C_T\Delta x}  \ (j\ge1),\\
\dot{\bar Q}_0&=-\frac{\Psi_0-\Psi_1}{L_T\Delta x},\\
\dot\Psi_0&=\frac{\bar Q_0}{C'_c}.
}
From these equations, by taking the continuous limit, $\Delta x\rightarrow 0$, $\Phi_j\rightarrow\Phi(x)=\Phi(j\Delta x)$, and $\Psi_j\rightarrow\Psi(x)=\Psi(j\Delta x)$, we obtain the wave equations for $x>0$ as
\eq{
\ddot\Phi(x)&=\frac{1}{C_TL_T}\pd{2}{\Phi(x)}{x},\\
\ddot\Psi(x)&=\frac{1}{C_TL_T}\pd{2}{\Phi(x)}{x},
}
with boundary conditions
\eq{
0&=-\frac{\Phi(+0)}{L_c}+\frac{1}{L_T}\pd{1}{\Phi(x)}{x}\Big|_{x=+0},  \label{BC}\\
\ddot\Psi(+0)&=\frac{1}{C'_cL_T}\pd{1}{\Psi(x)}{x}\Big|_{x=+0}.
}

Let us consider the inductively-coupled waveguide modes.
The eigenfunction with frequency $\omega$ is given by
\eq{
f_\omega(x)=\sqrt{\frac{2}{L}}\frac{ \cos\kakko{\frac{\omega}{v}x} + \zeta_I(\omega) \sin\kakko{\frac{\omega}{v}x} }{\sqrt{1+\zeta_I(\omega)^2}},
}
where $L$ is the length of the waveguide, $v=1/\sqrt{L_TC_T}$ is the speed of microwaves in the waveguide, and $\zeta(\omega)=\frac{L_Tv}{L_C\omega}$ is a dimensionless quantity defining the phase of the eigenfunction.
Then the positive frequency part of the flux variable $\Phi(x)$ is quantized as
\eq{
\Phi(x)^+=\sum_{k}\sqrt{\frac{\hbar Z_Tv}{2\omega_k}}f_{\omega_k}(x) b_k,
}
where $Z_T=\sqrt{L_T/C_T}$ is the characteristic impedance, $b_k$ is the annihilation operator of the eigenmode $f_{\omega_k}(x)$, and $\omega_k=\frac{\pi kv}{L}$.
Therefore, the flux variables $\Phi_0$ and $\phi$ are expressed in terms of the creation and annihilation operators as 
\eq{
\Phi_0 &= \sum_k\frac{\hbar Z_Tv}{\omega_kL} \frac{b_k+b_k^\dagger}{\sqrt{1+\zeta(\omega_k)^2}}\rightarrow \int_0^\infty \intd\omega \  \sqrt{\frac{\hbar Z_T}{\pi\omega}} \frac{b(\omega)+b(\omega)^\dagger}{\sqrt{1+\zeta(\omega)^2}},\label{Phi0}\\
\phi &=\sqrt{\frac{\hbar Z_R}{2}}(a+a^\dagger)    ,\label{phi}
}
where the limit $L\rightarrow\infty$ is taken in the second line.
Here, $a$ ($a^\dagger$) is the annihilation (creation) operator of microwave photons in the resonator.
The characteristic impedance for the resonator $Z_R$ is defined as $Z_R=\sqrt{L_R'/C_R}$.
By substituting Eqs.~\eref{Phi0} and~\eref{phi} to the inductive coupling term $-\phi\Phi_0/L_c$, we obtain 
\eq{
-\frac{\phi\Phi_0}{L_c}
=-\int_0^\infty \intd\omega\  \xi_0^I\sqrt{\frac{\omega}{1+(\omega/\omega_{\rm cutoff}^I)^2}} (a+a^\dagger)(b(\omega)+b(\omega)^\dagger),
}
where
\eq{
\xi_0^I&= \hbar\sqrt{\frac{1}{2\pi }\cdot\frac{Z_R}{Z_T}}, \\
\omega_{\rm cutoff}^I&=\frac{Z_T}{L_c}.\label{cutoff_I}
}

In a similar manner, we obtain the capacitive coupling term as
\eq{
\frac{q\bar Q_0}{C_R} 
=\int_0^\infty \intd\omega\  \xi_0^C\sqrt{\frac{\omega}{1+(\omega/\omega_{\rm cutoff}^C)^2}} (a-a^\dagger)(d(\omega)-d(\omega)^\dagger),
}
where
\eq{
\xi_0^C&= \hbar\sqrt{\frac{1}{2\pi }\cdot\frac{Z_TC_c'^2}{Z_RC_R^2}}, \\
\omega_{\rm cutoff}^C&=\frac{1}{Z_TC'_c},  \label{cutoff_C}
}
and  $d(\omega)$ is the annihilation operator of the capacitively-coupled waveguide mode with frequency $\omega$.

\begin{figure}[htbp]
\begin{center}
\includegraphics[width =0.7 \columnwidth]{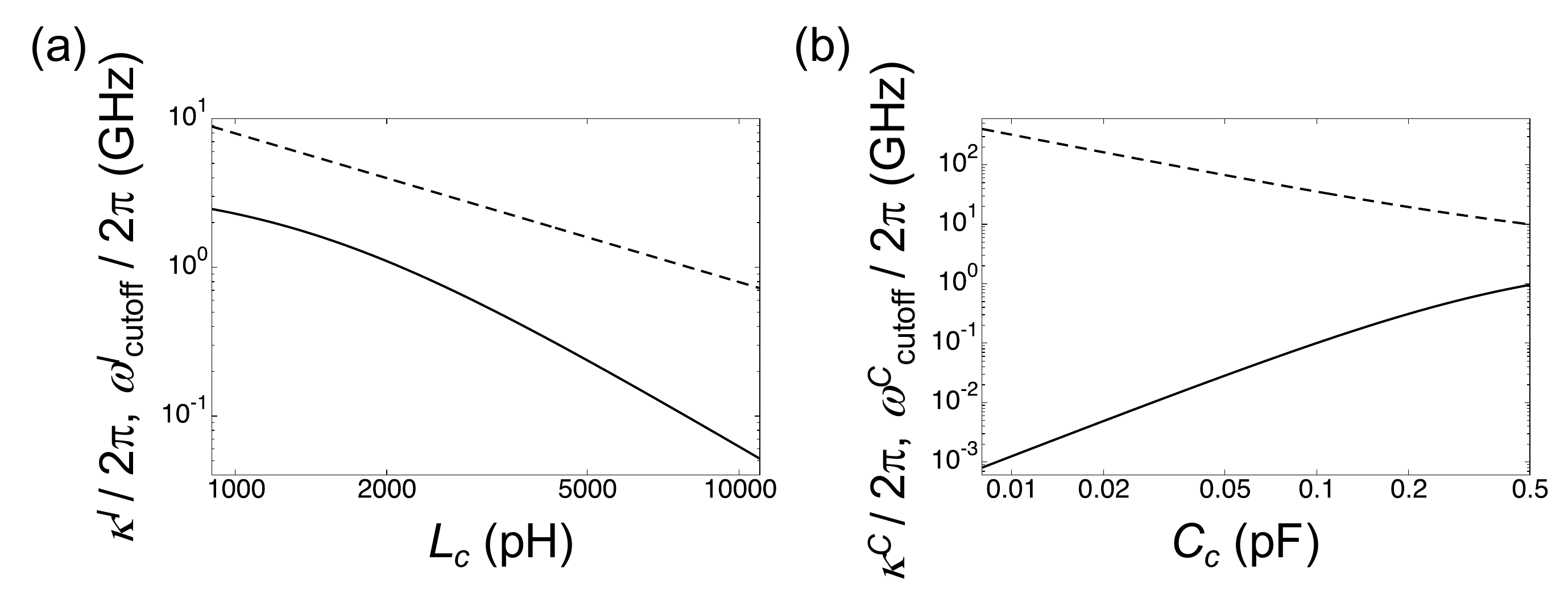}
\caption{Values of $\kappa$ (solid line) and $\omega_{\rm cutoff}$ (dashed line) plotted against (a) the coupling inductance ($L_c$) and (b) the coupling capacitance ($C_c$).
\label{parameters}}
\end{center}
\end{figure}

In the numerical calculation, the parameters are set to be $\omega_{\rm r}/2\pi=6$ GHz, $\Delta/2\pi=1.2$ GHz, $Z_W=50\ \Omega$, and $Z_R=30\ \Omega$.
In Fig.~\ref{parameters}, the values of $\kappa$ and $\omega_{\rm cutoff}$ are plotted as a function of the coupling inductance or capacitance in each case.
Since the left waveguide is directly connected to the Q-R system through the coupling inductance $L_c$, $\kappa^I$ takes a large value unless $L_c$ is sufficiently large, and is a decreasing function of $L_c$.
This property is special to the circuit in Fig.~\ref{both circuit}, and different from the circuits realized in Ref.~\cite{Yoshihara2017a,Yoshihara2018}, where the waveguide is coupled to the resonator through mutual inductance.

\section{Relation to the spin-boson model} \label{sec: spin-boson model}
In this section, we show that the Hamiltonian describing the circuit in Fig.~\ref{Circuit} (b) can be effectively mapped to the spin-boson model~\cite{Leggett1987}.
The total Hamiltonian is given by
\eq{
H_{\rm tot}=H_{\rm S} + H_{\rm E} + H_{\rm SE},
}
where
\eq{
H_{\rm S}&=\omega_{\rm r}a^\dagger a + \frac{\Delta}{2}\sigma_x + g\sigma_z (a+a^\dagger), \\
H_{\rm SE}&=\sum_k\xi_k(a+a^\dagger)(b_k+b_k^\dagger),\\
H_{\rm E}&=\sum_k\omega_kb_k^\dagger b_k.
}
To map the total Hamiltonian to the spin-boson model, we project the state space of the Q-R system onto the space spanned by the two low-lying energy states, $\{ \ket{\phi_0^{(-)}(\alpha)},\ket{\phi_0^{(+)}(\alpha)} \}$).
Let $\bar\sigma_j$ $(j=x,y,z)$ be the Pauli matrix in the basis of $\{\ket\uparrow\ket{-\alpha},\ket\downarrow\ket\alpha\}$.
Then the truncated Hamiltonian becomes the spin-boson model,
\eq{
 \frac{\Delta \ex^{-2\alpha^2}}{2}\bar\sigma_x + \sum_k\xi_k \bar\sigma_z (b_k+b_k^\dagger) +\sum_k\omega_kb_k^\dagger b_k.
}
Here, the ``localized state" corresponds to $\ket\uparrow\ket{-\alpha}$ and $\ket\downarrow\ket\alpha$.

The spectral function of this truncated Hamiltonian can be calculate as
\eq{
J(\omega)&=\frac{\pi}{2}\sum_k 32{\rm Re}[\alpha]^2\xi_k^2 \delta(\omega-\omega_k)   \nonumber\\
&=16\pi{\rm Re}[\alpha]^2\xi_0^2\omega\frac{1}{1+(\omega/\omega_{\rm cutoff})^2},
}
which corresponds to the Ohmic case.

\section{Symmetry of the total Hamiltonian and the CVS} \label{sec: symmetry}
The total Hamiltonian is invariant under a parity transformation $a\leftrightarrow-a$, $b_k\leftrightarrow-b_k$, and $\ket\uparrow\leftrightarrow\ket\downarrow$.
The ground state is expected to have the same symmetry, so that
\eq{
&c_0\ket\uparrow\otimes\ket{\alpha;\{\beta_j\} }+ c_1\ket\downarrow \otimes\ket{\alpha';\{\beta'_j\} }\nonumber\\
=&\ex^{i\theta}(c_0\ket\downarrow\otimes\ket{-\alpha;\{-\beta_j\} }+ c_1\ket\uparrow \otimes\ket{-\alpha';\{-\beta'_j\} }).
}
Then we have
\eq{
\alpha&=-\alpha',\\
\beta_j&=-\beta_j',\\
c_0&=\ex^{i\theta}c_1,\\
c_1&=\ex^{i\theta}c_0.
}
From the last two equations, we have $\ex^{i\theta}=\pm 1$ and $|c_0|=|c_1|=1/\sqrt 2$ from the normalization condition.
The choice of sign only affects the qubit energy term $\Delta\sigma_x$, and $\ex^{i\theta}=-1$ is chosen so that the qubit energy term reduces the total energy.

\section{Stationary state equations for CVS} \label{sec: stationary}
In this section, we derive the stationary state equations that the variational parameters of CVS should satisfy to minimize the total energy.
The total energy of the CVS is 
\eq{
E_{\rm CVS}&=& \braket{{\psi_C(\alpha,\{\beta_k\})}| H_{\rm tot}|{\psi_C(\alpha,\{\beta_k\})}}  \nonumber\\
&=&\omega_{\rm r}|\alpha|^2 - g(\alpha+\alpha^*) + \sum_k \omega_k |\beta_k|^2 \nonumber\\
&&\pm \sum_k\xi_k(\alpha\pm\alpha^*)(\beta_k\pm\beta_k^*)  - \frac{\Delta}{2}\expo{-2(|\alpha|^2 +\sum_k|\beta_k|^2)},
}
depending on whether the R-W coupling is inductive or capacitive.
The variational parameters minimizing the total energy should satisfy $\partial E_{\rm CVS} /\partial \alpha^*=0$ and $\partial E_{\rm CVS} /\partial\beta_k^*=0,\forall k$, which read
\eq{
\omega_{\rm r}\alpha + \Delta\alpha\ex^{-2(\alpha^*\alpha+\sum_k\beta_k^*\beta_k)}- g + \sum_k\xi_k(\beta_k\pm\beta_k^*)&=&0,  \label{stat alpha}\\
\omega_k\beta_k +  \Delta\beta_k\ex^{-2(\alpha^*\alpha+\sum_k\beta_k^*\beta_k)} + \xi_k(\alpha\pm\alpha^*)&=&0,\forall k.  \label{stat beta}
}
From Eq.~\eref{stat beta}, $\beta_k$ can be expressed in terms of $\alpha$ and the collective variable $S=\sum_k\beta_k^*\beta_k$ as follows:
\eq{
\beta_k=-\frac{\xi_k(\alpha\pm\alpha^*)}{\omega_k+\Delta\ex^{-2(\alpha^*\alpha+S)}}.
}

In the inductive coupling case, we have
\eq{
\beta_k=-\frac{2\xi_k {\rm Re}[\alpha] }{\omega_k+\Delta\ex^{-2(\alpha^*\alpha+S)}},\label{beta ind}
}
which is real. Substituting Eq.~\eref{beta ind} into Eq.~\eref{stat alpha} and the definition of $S$, we obtain
\eq{
\omega_{\rm r}\alpha + \Delta\alpha\ex^{-2(\alpha^*\alpha+S)}- g - \sum_k 4{\rm Re}[\alpha]f_1(\Delta\ex^{-2(\alpha^*\alpha+S)})&=&0, \label{ind stat1} \\
4{\rm Re}[\alpha]^2f_2(\Delta\ex^{-2(\alpha^*\alpha+S)})&=&S,
}
where we have defined
\eq{
f_1(x)&=&\sum_k \frac{\xi_k^2}{x+\omega_k}, \\
f_2(x)&=&\sum_k \frac{\xi_k^2}{(x+\omega_k)^2}.
}
From Eq.~\eref{ind stat1}, we see that $\alpha$ is also real.
Although there are originally an infinitely large number of parameters $\{ \beta_k\}$, we obtain a set of  closed equations for $\alpha$ and $S$, and once we obtain the stationary solution for $\bar{\alpha}$ and $\bar{S}$, $\beta_k$ can be obtained from Eq.~\eref{beta ind}.
In the continuous limit of $L \rightarrow \infty$, $f_1(x)$ and  $f_2(x)$ can be explicitly calculated as follows:
\eq{
\fl f_1(x)\rightarrow
&\frac{\xi_0^2\omega_{\rm cutoff}^2}{x^2+\omega_{\rm cutoff}}
\left(  -x\log\frac{\omega_{\rm cutoff}}{x} + \frac{\pi\omega_{\rm cutoff}}{2} \right), \\
\fl f_2(x) \rightarrow
&\xi_0^2\omega_{\rm cutoff}^2 \left(  -\frac{1}{x^2 + \omega_{\rm cutoff}^2} +\frac{x^2-\omega_{\rm cutoff}^2}{(x^2+\omega_{\rm cutoff}^2)^2}  \log\frac{x}{\omega_{\rm cutoff}} +   \frac{\pi\omega_{\rm cutoff}x}{(x^2+\omega_{\rm cutoff}^2)^2}  \right).
}

In the capacitive coupling case, we have 
\eq{
\beta_k=-\frac{2i\xi_k {\rm Im}[\alpha]}{\omega_k+\Delta\ex^{-2(\alpha^*\alpha+S)}},\label{beta cap}
}
which is purely imaginary.
Substituting Eq.~\eref{beta ind} into Eq.~\eref{stat alpha}, we obtain
\eq{
\omega_{\rm r}\alpha + \Delta\alpha\ex^{-2(\alpha^*\alpha+S)}- g + i\sum_k 4{\rm Im}[\alpha]f_1(\Delta\ex^{-2(\alpha^*\alpha+S)})&=&0. \label{cap stat1}
}
Subtracting Eq.~\eref{cap stat1} from its complex conjugate, we obtain
\eq{
(\omega_k+\Delta\ex^{-2(\alpha^*\alpha+S)})(\alpha-\alpha^*)=0,
}
so that $\alpha$ is real.
Since ${\rm Im}[\alpha]=0$, we have $\beta_k=0$ from Eq.~\eref{beta cap} and hence $S=0$, and finally we obtain a closed equation for $\alpha$ as
\eq{
(\omega_{\rm r} +\Delta \ex^{-2\alpha^2})\alpha=g.  \label{cap stat}
}
Equation~\eref{cap stat} is the same as the stationary state equation for the closed quantum Rabi model, i.e., $\xi_k=0$.
This result means that the CVS does not provide a good approximation for the qubit-resonator-waveguide ground state in this case.

\section{Validity of CVS in inductive coupling}  \label{sec: validity}
\begin{figure}[htbp]
\begin{center}
\includegraphics[width = 0.8\columnwidth]{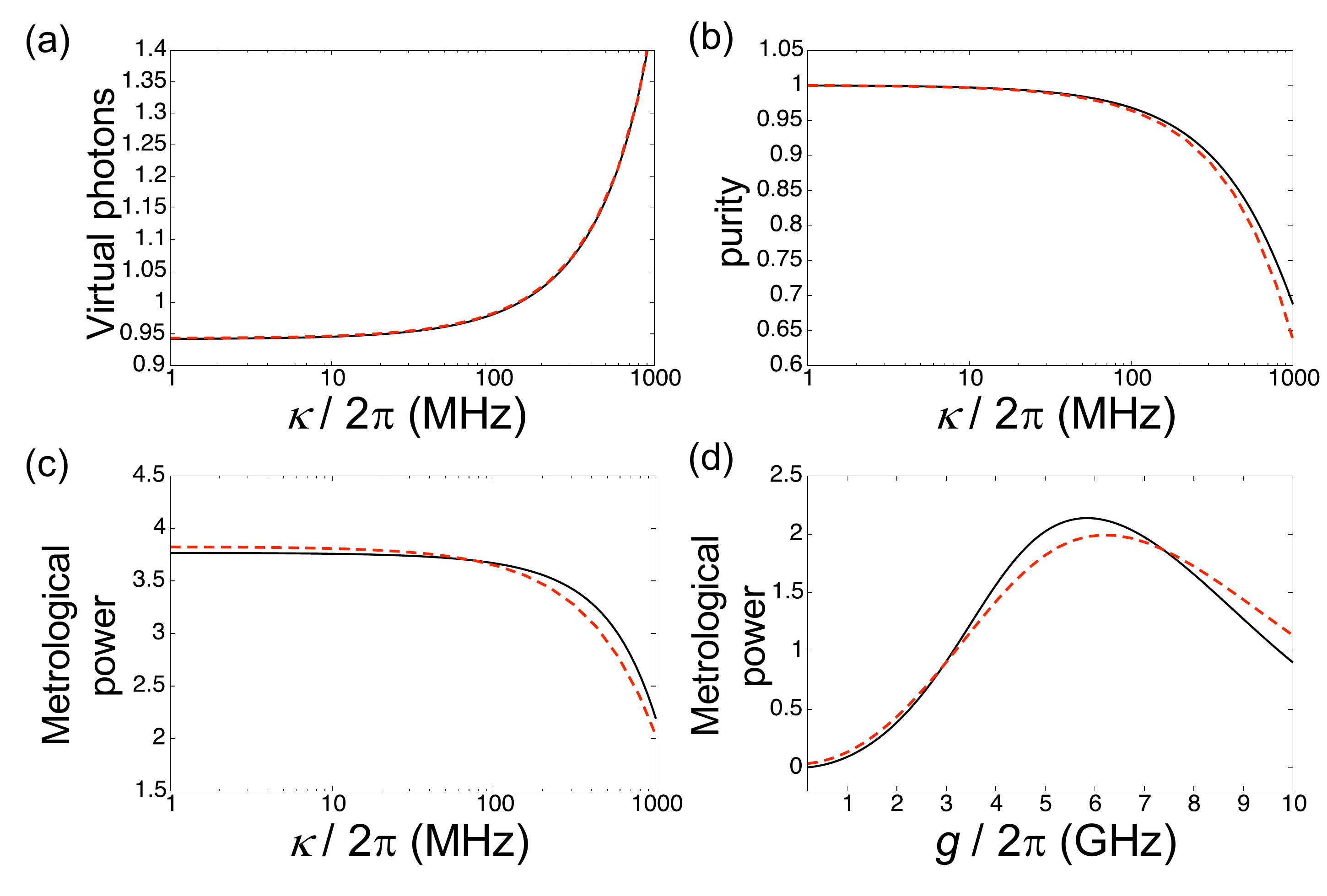}
\caption{The average virtual photon number (a), the purity (b), and the average metrological power, plotted against the bare loss rate $\kappa$.
(d) The average metrological power plotted against the Q-R coupling strength $g$.
All the quantities are calculated from the CVS (black solid line) and the numerical diagonalization (red dashed line).
The loss rate is set to $\kappa/2\pi=$ 1000 MHz, and the Q-R coupling is set to $g/2\pi=$ 6 GHz.
\label{Validity}}
\end{center}
\end{figure}

In this appendix, we confirm the validity of the CVS in the inductive coupling case, by comparing the result from the CVS and the numerical diagonalization for a few waveguide mode case.
We see that in Fig.~\ref{Validity} (a) and (b), the number of virtual photons and the purity calculated from the CVS and the numerical diagonalization agree quantitatively.
We also see that from Fig.~\ref{Validity} (c) and (d), the nonclassical property of the system, measured by the metrological power can also be described by the CVS.

\section*{References}  
\bibliography{library}

\end{document}